# "Blockchain-Enabled Zero Trust Framework for Securing FinTech Ecosystems Against Insider Threats and Cyber Attacks"

**Avinash Singh**

*Department of Computer Science and Information Technology Mahatma Gandhi Central University, East-Champaran, Bihar, India.*

**(Corresponding author's email : mgcu2019csit6002@mgcub.ac.in)**

**Prof. Vikas Pareek**

*Department of Computer Science and Information Technology*
*Mahatma Gandhi Central University, East-Champaran, Bihar, India.*
*vikaspareek@mgcub.ac.in*

**Dr. Asish Sharma**

*Department of Computer Engineering,*
*Manipal University, Jaipur*
*ashishmudgal.mnit@gmail.com*

**Abstract**:

Fintech provides technological services to increase operational efficiency in financial institutions, but traditional perimeter-based defense mechanisms are insufficient against evolving cyber threats like insider attacks, malware intrusions, and Advanced Persistent Threats (APTs). These vulnerabilities expose Fintech organizations to significant risks, including financial losses and data breaches. To address these challenges, this paper proposes a blockchain-integrated Zero Trust framework, adhering to the principle of "Never Trust, Always Verify." The framework uses Ethereum smart contracts to enforce Multi Factor Authentication (MFA), Role-Based Access Control (RBAC), and Just-In-Time (JIT) access privileges, effectively mitigating credential theft and insider threats, the effect of malware and APT attacks.

The proposed solution transforms blockchain into a Policy Engine (PE) and Policy Enforcement Point (PEP), and policy storage, ensuring immutable access control and micro-segmentation. A decentralized application (DApp) prototype was developed and tested using STRIDE threat modeling, demonstrating resilience against spoofing, tampering, and privilege escalation. Comparative analysis with Perimeter-based systems revealed a trade-off: while the framework introduced a marginal latency increase (74.0 ms vs. 49.33 ms) and reduced throughput (30.77 vs. 50.0 requests/sec), it significantly enhanced security by eliminating single points of failure and enabling tamper-proof audit trails.

Experimental validation on a 200-node simulated network confirmed the framework's robustness, with future optimizations targeting Layer-2 solutions for scalability. This work bridges the gap between Zero Trust theory and practical blockchain implementation, offering Fintech organizations a decentralized, cost-effective security model.

**Introduction**

Fintech leverages technology to enhance the efficiency and productivity of various financial activities within a financial organization. Financial institutions like banks, insurance companies, money lending agencies, etc., are essential components of the fintech ecosystem. The other elements are government, financial customers, and Fintech startups. These different components synchronize themselves with technological threads to form a Fintech ecosystem. This section will explore the global growth of Fintech startups to understand the significance of these companies and the advancement of technology in financial operations. The Statista Fintech report forecasts startup growth from 2018 to 2023. The forecasted numbers of Fintech startups in the year 2023 are 11651 in America, 9,681 in EMEA, and 5,061 in APAC countries [1]. However, alongside the growth in technology, there has been an increase in cyber threats. Hackers are leveraging advanced technology to gain unauthorized access to confidential information, such as user credentials and data breaches, impacting the fintech industry. We examined various cyber-attacks between 2020 and 2022, and Figure 1 illustrates their effects on fintech institutions. Examining the list of attacks worldwide published by Carnegie [2] yields valuable insights. The figure illustrates the different categories of attacks and their counting in numbers on the fintech organizations.

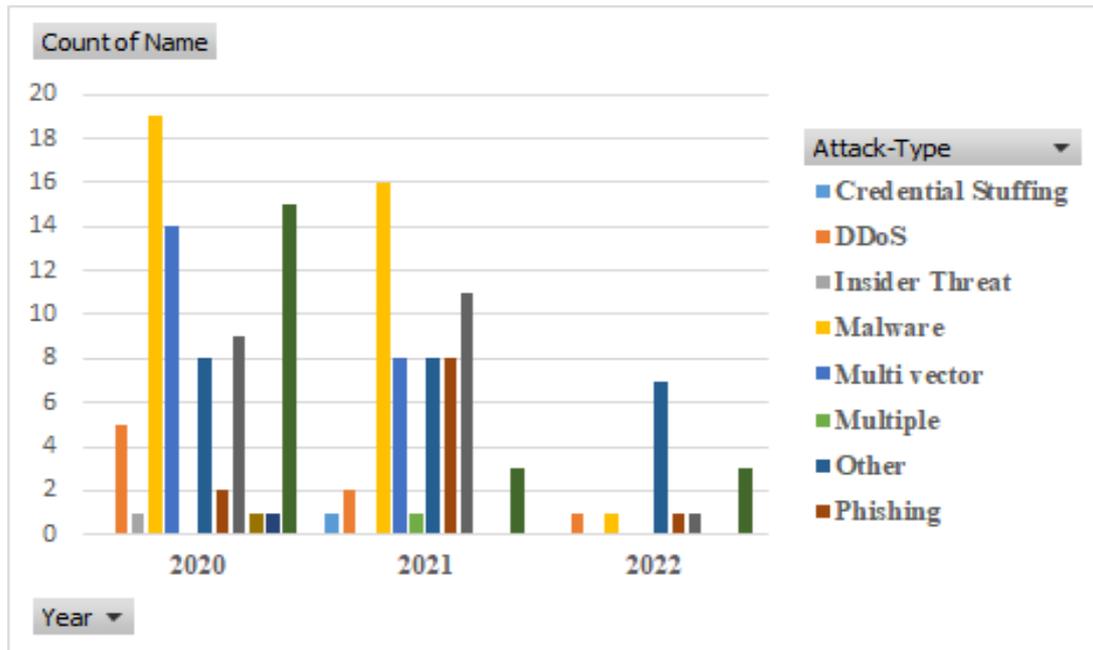

**Figure 1**: Number of attacks on Fin-Tech Organizations from 2020 to 2022.

The analysis of the attack pattern on the Fintech companies predicts different kinds of attack categories, including credential stuffing, malware attacks performing espionage, DDoS attacks, insider threats and multi-vector threats. These threats lead to data breaches, operational disruption, and theft of credentials and sensitive information in fintech organizations. Malware attacks can facilitate a more advanced threat known as Advanced Persistent Threats (APTs) that extract sensitive information over an extended time without being observed [3]. Insider threats represent another significant category, with approximately 82% of insider incidents involving the exploitation of legitimate access, making it a considerable concern for information security [4].

The root cause of the aforementioned security breaches is the inadequacy of traditional Perimeter-based security defense strategies, which fail to sufficiently safeguard financial institutions against evolving threats [5][6]. Incorporating cutting-edge technologies introduces substantial challenges for security professionals as adversaries leverage sophisticated artificial intelligence techniques to compromise Fintech enterprises. Organizations implement various protective measures in a traditional perimeter-based security model through policies specifically designed for their network resources [7].

Perimeter-based security divides the Fintech ecosystem into discrete subsystems, such as intranet and extranet. Security mechanisms create boundaries and operate, assuming that resources within these boundaries are secure. Nonetheless, the obsolescence of these security devices, software, and policies renders the Perimeter-based security model insufficient [8],[9].

To overcome the shortcomings of Perimeter-based security, which is still creating feasible attack surfaces for the attackers, J. Kindervag introduced the concept of Zero Trust, whose key motto is "Never Trust, Always Verify [10]. The organization verifies devices, users, applications, and processes before granting access to its resources. Any technology that meets the key tenets of Zero Trust can be used to implement its principles. The key tenets of Zero Trust include verifying the identity of the user, location, device attributes, and other factors. Enforcing least-privileged access minimizes resource uses through Just-In-Time privilege escalation, risk-based adaptive control, data protection, and network segmentation to reduce the scope of breaches and prevent lateral movement and damage [8]. This approach reduces the likelihood of attacks that typically occur under a perimeter-based security model. This article proposes a framework that combines the ZeroTrust principle with blockchain technology to enhance the security posture of Fintech organizations against internal threats, including malware attacks and APTs. The proposed framework leverages the Ethereum blockchain and smart contracts to implement Zero Trust security.

The following are the significant contributions of the proposed work:

1. **A Zero Trust security framework:** This framework enhances the security of FinTech organizations and protects their operations from various threats, including insider threats and credential theft caused by malware.

2. **Efficient blockchain integration**: Multifactor authentication and role-based access control, including Just-In-Time access control, are implemented using the Ethereum blockchain to enforce the Zero Trust model for FinTech organizations. This approach will reduce operational costs, as blockchain is an open-source technology.

3. **Experimental validation:** The framework is evaluated using the STRIDE threat model and performance metrics such as latency and throughput. A comparative analysis with a perimeter-based security model is conducted in a simulated 200-node network to measure attack resilience and authentication efficiency. FinTech organizations can use the outcome of the proposed framework to develop **DApp (**Decentralized Applications) for FinTech operations. Our results validate the robustness of the Zero Trust framework comparing the perimeter-based security model.

4. The remainder of this article is structured as follows: Section 2 presents a state-of-the-art literature survey. Section 3 covers preliminaries, including an overview of the Zero Trust model of security and its architecture. Section 4 discusses blockchain and its architecture, including smart contracts and procedures for developing DApps, and its role in supporting the Zero Trust paradigm Section 5 outlines the proposed framework along with a use case example. Section 6 presents a detailed description of the smart contracts used and discussion of findings. Section 7 concludes the article and outlines the future directions.

**Related work:**

Blockchain technology combined with Zero Trust Architecture (ZTA) establishes a robust security framework that mitigates, ensuring no entity is trusted by default and thoroughly verifying every access request. This approach leverages blockchain's inherent properties of decentralization, immutability, and transparency to enhance Zero Trust principles by addressing authentication, access control and data integrity challenges. In the FinTech sector, this convergence strengthens security mechanisms, protects sensitive financial transactions and reduces fraud. The following sections explore how blockchain can augment zero trust principles across various domains, including fintech security, highlighting key implementations and benefits. This study examines the intersection of Zero Trust, blockchain, and fintech security. The literature review is categorized into three themes: (i) the role of Zero Trust in fintech security, (ii) blockchain's contributions to cybersecurity, and (iii) strategies for integrating blockchain with the Zero Trust framework to enhance resilience against threats.

The following section presents a comprehensive study of the state-of-the-art literature:

Guo, Shaoyong, et al. [11] analyzed how blockchain enhances authentication and data sharing in IoT environments, particularly Zero Trust settings. Blockchain uses smart contracts and consensus mechanisms to ensure data privacy and trustworthiness while preventing unauthorized data sharing. This approach is particularly beneficial in edge computing, which supports dynamic identity authentication and fine-grained access control, enhancing security and efficiency. However, the study identifies research gaps, including the need for improved interoperability among different IoT platforms, stronger security measures, and a more efficient authentication mechanism that can operate effectively in a decentralized environment.

D. et al. [12], in their article titled "Zero Trust in Edge Computing Environment: A Blockchain-based Practical Scheme," proposed a blockchain-driven Zero Trust architecture to improve the security of edge computing for smart cities, the authors argued that the traditional security measures are not more effective due to dynamic nature of IoT devices, and hence used the zero trust model that provides fine-grained access control, continuous verification, a distributed authentication algorithm and the RAFT consensus algorithm for secure identity verification. However, a key limitation of their work is the lower transactions per second (TPS) compared to pure blockchain platforms. The research gap identified involves the practical implementation of Zero Trust in edge computing, suggesting a need for further optimization.

Daah, C. et al. [13], in their article "Enhancing Zero Trust Models in the Financial Industry through Blockchain Integration: A Proposed Framework," introduced a Zero Trust framework integrated with blockchain designed for the financial industry to address the growing complexity of threats. The authors developed a prototype banking application to test the framework's effectiveness through vulnerability scanning and performance testing. The solution enhances the security against both internal and external threats. However, a key research gap is the framework's limited acceptability to the other sectors, suggesting a need for border validation.

Susanto H. et al. [14] examined the effectiveness of blockchain technology in mitigating cyber threats faced by the fintech industry. As an immutable ledger and decentralized ledger, blockchain enhances security by encrypting and securely storing transaction data. This approach significantly reduces risk associated with data leakage and misuse, as these are serious concerns faced by the fintech industry and affect financial services and stakeholders. No specific research gap is mentioned in the article.

Din, Ikram Ud, et al. [15], in their paper "Securing the Metaverse: A Blockchain-Enabled Zero-Trust Architecture for Virtual Environments, "propose a blockchain-integrated Zero Trust framework to enhance the cybersecurity in network infrastructure. It shows that the blockchain-enabled Zero Trust architecture is much better than conventional security systems and has better threat detection and transaction ability. The research consists of simulations to compare performance metrics like intrusion detection rates and breach response times. The architecture is suitable for virtual environments like Metaverse. The study acknowledges the limitations of the proposed work in a real-world environment and the need for testing and validation in real-world scenarios. The research also identifies a critical gap in practical implementation of such framework, calling for future work to bridge theory and deployment.

Zaabar, B. et al.[16] their paper "Health Block: A Secure Blockchain-based Healthcare Data Management System" address a critical need for enhanced privacy and security in Electronic Health Records (EHRs) due to vulnerabilities present in centralized databases, which are prone to cyberattacks. The authors proposed a decentralized architecture using OrbitDB and IPFS for secure record storage, while a Hyperledger Fabric-based blockchain network to store data hashes and enforces access control during retrieval. Performance evaluations confirm the system's robustness and efficiency in improving security and data integrity for healthcare management. However, the study acknowledges scalability limitations and a research gap in scaling the solution and integrating advanced security features.

Chaudhry et al. [17], in the paper titled "Zero Trust-Based Security model against data breaches in the banking sector: A blockchain consensus algorithm," investigates the need for a robust security model in the banking system due to rising cyber threats. The study proposes integrating Zero Trust security principle with blockchain technology to strengthen system security. Through literature analysis and iterative testing to evaluate, the authors evaluate their framework's feasibility, which leverages blockchain's immutability and decentralization to secure the transactions. However, the study identifies the limitations in real-time implantation and blockchain adoption due to technical constraints. The study further explores the research gap associated with the integration of blockchain and Zero Trust model for banking, which needs further exploration for practical implementation in real-world scenarios.

Rivera, J. et al. [18], in the paper "Secure Enrollment Token Delivery Mechanism for Zero Trust Networks Using Blockchain," introduce a method for the delivery of secure enrollment tokens for Zero Trust networks using blockchain technology, the technique focused on the integration of JSON web tokens and Non-Fungible Tokens (NFT) to increase the security during the enrollment process. The method includes encrypting the JWT with the owner's public key, ensuring secure ownership and distribution of One-Time Tokens (OTT). The application of the proposed work includes enhancing identity verification in the Zero Trust framework and using the permissioned Ethereum blockchain for secure enrollment. The limitations include high transaction fees associated with public blockchain networks. The identified research gap is a more detailed exploration of token standards with practical applications and uses of tokens in the authentication process.

**Table 1: Summary of research insights on integrating Blockchain with ZTA**

| Ref. No. | Insights | Methods Used | Applications | Limitations | Research Gap |
|---|---|---|---|---|---|
| Patil et al. [19] | A consensus algorithm consisting of blockchain and Zero Trust model is created, which is helpful for the students and enhances data security while preventing the malicious node from gaining access. | The paper proposes using a consensus algorithm to build a Zero Trust Model, ensuring that each node is responsible for approving transactions before they are committed. | Creation of a college placement system, functional for students to maintain an immutable certificate vault based on Zero Trust principle and blockchain. | Malicious nodes in the network may attempt to disrupt the consensus algorithm by causing transactions to be aborted, potentially impacting the overall system reliability and performance. | The paper does not explicitly mention any research gaps. Future research will be based on the proposed algorithm's real-time use, including the scalability of the system. |
| Wylde [20] | Blockchain technology can enhance Zero Trust as it is decentralized, and all the operations are validated in a distributed manner. Using smart contracts will decrease the vulnerabilities associated with traditional trust models. | The paper discusses the concept of Zero Trust as being based on no presumptive trust and a risk-based approach, emphasizing the continuous verification of trust. The development of a new framework for the implementation of the trust | (ZTA) is highlighted as a cybersecurity plan of an enterprise that incorporates Zero Trust concepts, encompassing component relationships, workflow planning, and access policies. | The paper acknowledges that it has limitations regarding scope and exploration being a short one in nature, including that there is more depth and breadth to be explored in the concept of Zero Trust and Zero Trust architectures. | The paper highlights a research gap in the scholarly trust and organization studies literature regarding applying prominent trust formation models, such as the Integrative Trust Model (ITM), in the |
| | | principle | | | context of Zero Trust and Zero Trust Architectures. |
| Alevizos et al.[21] | Blockchain ensures immutable logs, enhancing endpoint security. Detects log changes, bolstering defence against APTs. Smart contracts enable automated, real-time anomaly responses. | Integrates ZTA with blockchain to boost endpoint security. Focus on preventing lateral movement and APTs. Provides an overview of existing Zero Trust Architecture models and their practical implementations. Highlights ZTA's role in securing endpoints. | ZTA defends against lateral movement in borderless networks by securing endpoints. Blockchain integration enhances intrusion detection, leveraging immutability against APTs. | APTs effectively bypass intrusion detection, remaining stealthy. ZTA's vulnerability lies in compromised, authenticated endpoints. | Challenges exist in integrating ZTA with blockchain to detect Advanced Persistent Threats (APTs). APTs often bypass current intrusion detection systems. Further research is needed to develop robust countermeasures. |
| Li et al. [22] | Utilizes blockchain for Zero Trust in edge computing. Proposes digital identity for two-way authentication and fine-grained access. Enhances traceability with identity data on a distributed ledger. Uses improved RAFT for multi-party consensus in smart city security. | The Digital Identity Model for Edge computing establishes two-way authentication for edge nodes and terminals. Enables fine-grained authorization and access control. Uses improved RAFT for multi-party consensus and trust quantification. | Applies blockchain-based ZTA in edge computing for smart cities. Addresses fuzzy boundary security and dynamic identity authentication. Proposes a digital identity model for fine-grained access control. | Traditional protocols rely on a trust authority, conflicting with Zero Trust principles. The proposed protocol categorizes devices as trusted or suspected. Only trusted and suspected devices can authenticate, limiting network participation. | Does not assess network conditions' effects on Zero Trust and blockchain solutions. Lacks scalability and performance analysis for many edge nodes. |
| Liu et al. [23] | Zero Trust in Fintech enhances security by treating all transactions as untrustworthy, requiring continuous verification. It reduces the risks of data breaches and fraud through | The paper proposes a blockchain solution for Zero Trust IoT to ensure anonymity, authentication, privacy, trustworthiness, and fairness. It uses smart contracts, | Decentralized info-sharing in IoT without Trusted Third Party. Uses blockchain and smart contracts. The reputation chain records transactions and ranks participants by | Central servers may be compromised, risking decentralized components. The protocol focuses on replay attack resilience. The research does not fully address other | Combines decentralization, privacy, fairness, and security in Zero Trust environments, and lacks comprehensive support A proposed blockchain protocol |

| | | | | |
|---|---|---|---|---|
| constant identity and device checks. Blockchain aids in achieving authentication, privacy, and fairness in data sharing. This approach supports the need for strong security in digital financial systems. | voting, and consensus to filter out fabricated information and prevent unauthenticated data sharing. | reputation. | potential attack vectors and constraints. Unexamined vulnerabilities could impact real-world robustness. | aims to operate without a trusted third party, addressing literature gaps on autonomous information-sharing in adversarial settings. |

A review of previous literature indicates that blockchain and Zero Trust have been implemented across various domains, including IoT, edge computing, identity management, and fintech, to enhance security and mitigate the threats. The literature survey categorizes prior work into two main areas: (1) theoretical frameworks designed to secure the existing system [16],[19], [20], and (2) practical implementations of the Zero Trust paradigm in IoT [23], cloud, and edge computing [22].

Several studies have explored the integration of fintech, Zero Trust and blockchain [12],[13], [14], primarily focusing on application-centric approaches to counter cyber threats such as DoS attacks, Man-in-the-Middle attacks, and other security breaches. While these works contribute to enhancing fintech security, they do not evaluate blockchain's behaviour as a Zero Trust enforcement mechanism. Specifically, they lack an analysis of blockchain as a Zero Trust policy engine (PE) and Policy Enforcement Point (PEP)and policy storage.

The upcoming section "Architecture of the Zero-Trust Security model, "will address this gap by presenting a framework in which blockchain functions as a Zero Trust engine and enforcement mechanism, ensuring secure authentication, access control and policy enforcement within fintech systems.

The proposed framework uses blockchain and Zero Trust to mitigate internal attacks in fintech organizations, functioning as a blockchain-enabled Zero Trust enforcer. The framework will counter the attacks that are occurring due to the use of perimeter-based defence architecture. The framework involves the development of Decentralized Applications (DApps), built using smart contracts written in Solidity, a core component of the Ethereum blockchain. These Solidity-based smart contracts will implement Multi Factor Authentication (MFA), Role-Based Access Control (RBAC), and Just-In-Time (JIT) access control, ensuring a dynamic and adaptive security model within fintech systems.

To evaluate the framework's effectiveness, we will conduct a comparative analysis between the traditional perimeter-based security models and the proposed Zero Trust framework using the STRIDE threat modelling technique. Additionally, simulation techniques will be applied to compare the duo based on latency and throughput. The findings will demonstrate the robustness of the blockchain-driven Zero Trust approach in strengthening fintech security.

In the next section, we provide an overview of Zero Trust, Zero Trust Architecture (ZTA), Blockchain and Decentralized Applications (DApp) to establish a foundational understanding of our framework.

**Overview of the Zero Trust Security Model**:

The Zero Trust security model is an emerging security concept that challenges the traditional perimeter-based security approach. It is based on the fundamental principle of "Never Trust, Always Verify." The model assumes that threats exist both inside and outside the network. This model verifies every entity and associated process at the macro and micro levels before giving access to the ecosystem's resources. The zero-trust model is designed to secure modern enterprises by implying the following key tenets:

**i) Verify Explicitly:** Always authenticate and authorize access to the resources based on several factors (user identity, device information including location, behaviour and security posture). Use Multi-Factor Authentication (MFA) and risk-based access control mechanisms.

**ii) Least privileged Access:** Grant Users and systems only the minimal access necessary to perform their tasks. Use Just-in-Time (JIT) Access and or Attribute-Based

Access Control (ABAC) principles to minimize the exposure to sensitive resources of the ecosystem.

 **iii) Assume Breach:** Design a security strategy for the organization that assumes a breach will occur or has already occurred.

 **iv) Micro-Segmentation:** Divide the network into smaller subzones to limit the lateral movement of the attack around different nodes in the network. This can be done using hardware-driven segmentation or using Software-Defined Networking (SDN) techniques.

 **v) Continuous monitoring and Analytics:** Use real-time monitoring, AI-driven threat detection and behavioural analysis techniques to identify suspicious activities.

 **vi) Secure all endpoints and devices:** Ensure all the endpoint devices are updated by applying the patches.

 **vii) Secure Data and Encrypt Everything:** The data associated with the system, either in transit or in flow, should be encrypted.

 **viii) Automate security and Response:** Use AI and automation to enforce different security policies of the organization to detect threats and respond to the threats.

To implement all these tenets, an organization needs technologies to satisfy all the key principles of Zero Trust. Further, the organization is free to adopt any kind of technological implementation to achieve the key tenets of Zero Trust. Using different technologies should satisfy the essential requirement of the Zero Trust principles so that the organization has a better security status and can overcome different attack scenarios. Zero Trust principles can be implemented according to the organization's security posture and by analyzing the cost-benefit analysis [24]. Evaluating blockchain technology has given new insight and positive results in implementing Zero Trust principles, as it is also considered trustless [25][26].

**Architecture of the zero-trust security model**

 According to the NIST's documentation (NIST SP 800-207) of the Zero Trust security model, the architecture of the zero-trust model is divided into two main components. The first component is the Policy Enforcement Point (PEP), and the second one, the Policy Decision Point (PDP). Policy decision points include the components known as Policy Engine (PE) and policy administrator (PA). In addition, policy storage, resources, and users are part of the Zero Trust security model.

The Zero Trust policy enforcement point constitutes a vital layer between internal and external users and the various system resources that an organization manages. When a user initiates a request to access the organization's valuable resources and services, whether from within or outside the company, this request is routed to the PDP.

The PE within the PDP is responsible for verifying the security requirements defined and designed by the Policy Administrator (PA) of the organization in the form of security policies, which are stored in a designated policy storage in the form of predefined access control rules. The PE meticulously assesses the user's request against the established access control rules within policy storage. If the request aligns with the defined security requirements, access to the resources and services will be granted. Conversely, if the request does not meet the specified criteria, it will be promptly denied by the policy engine.

The security policies under consideration may encompass a range of factors, including the verified identity of the user, the identity of the device utilized, the geographical location from which the request originates, the overall health status of the device, and any additional factors that the fintech company may opt to integrate to implement a robust Zero Trust architecture effectively. The accompanying diagram in Figure 2 elucidates the intricate structure of the Zero Trust security model, showcasing its various components and their interactions within the broader security framework.

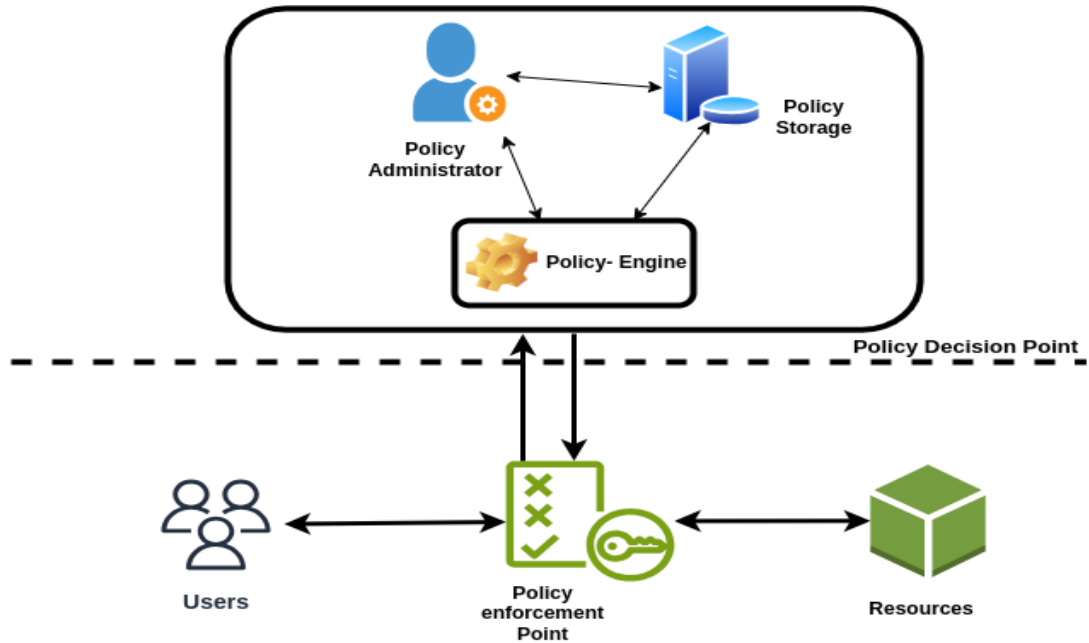

**Figure 2**: Zero Trust Security Model Architecture.

As discussed earlier, we are taking blockchain as a PE, policy storage, and PEP and considering it a Device Agent deployment model with zero trust [27]. The blockchain system stores and checks rules associated with the security requirements for a given fintech organization. The following section will delve into blockchain architecture and its various components to better understand blockchain technology.

**Introduction to Blockchain, Blockchain Architecture, overview of smart contract and Decentralized Application development process.**

**Blockchain**

Blockchain represents an innovative technology based on the concept of decentralization of data storage and security based on consensus protocols. Blockchain records transactions across many nodes, ensuring that the stored information on the distributed ledger is immutable and resistant to unauthorized alterations. This technology has gained attention in different domains, including cryptocurrencies, finance, healthcare data management, supply chain management and voting systems [28].

By enabling secure transactions, blockchain eliminates the need for intermediaries or central administration of data storage, reducing the cost of the system. Moreover, the inherent features of the blockchain, such as consensus mechanism and cryptographic security, create a more secure and transparent environment for data exchange. Due to the mentioned features, different organizations and governments continue to explore the usage of the blockchain in various domains, including security, data exchange and many more. The advancement of the blockchain also affects the institution's business and operational logic and is more effective than previous platforms used for these operations. In the next section, we will discuss blockchain architecture, insight into various categories of blockchains and the development process of Decentralized Applications (DApps) and consensus protocols.

**Blockchain and blockchain architecture**

The blockchain architecture consists of decentralized, interconnected nodes that share a common immutable ledger containing transaction data. The nodes are connected through the nodes by their global identity (public key) and perform the transactions (cryptocurrency-driven blockchain) or any set of computations (in the case of Ethereum blockchain) on the Ethereum virtual machine. The transactions or set of operations are validated through the consensus mechanisms, including Proof of Work (PoW), Proof of Stake (PoS), and Proof of Burn (PoB) and many more to attain a common understanding between all nodes to add the transaction or validate the operation in the chain and add the transaction or operation to the immutable blockchain [29] [30].

In our study, we are utilizing the Ethereum blockchain; therefore, the subsequent section will examine the Ethereum blockchain and its ecosystem, including the various types of nodes, the consensus mechanism, and the process of developing a decentralized application using the Ethereum blockchain.

**Ethereum blockchain**

The concept of blockchain was introduced by Satoshi Nakamoto in 2008, and it was implemented in 2009 as a core component of the cryptocurrency Bitcoin [31]. The Bitcoin blockchain is dedicated to performing cryptocurrency transactions based on mnemonics used for this purpose and, therefore, cannot convey other operations such as computation etc. To overcome the computational limitations of the Bitcoin blockchain, the Ethereum blockchain was introduced in 2015 by Vitalik Buterin [32].

Ethereum represents a significant advancement beyond Bitcoin, introducing the concept of smart contracts. These self-executing contracts are programmed to automatically enforce and execute the terms of an agreement when predefined conditions are met. Ethereum's innovative approach has enabled the development of Decentralized Applications (DApps) that leverage smart contracts to facilitate secure and transparent transactions without intermediaries [33].

Ethereum's blockchain architecture comprises interconnected nodes that validate transactions and execute smart contracts across the Ethereum Virtual Machine (EVM). The nodes in the Ethereum blockchain are divided as

**i) Full nodes:** Maintain a complete copy of the blockchain, validate the smart contracts and transactions and participate in the network consensus.

**ii) Light nodes:** Also known as thin clients, store only block headers and rely on the full nodes to retrieve the complete blockchain information.

**iii) Mining nodes:** These nodes are responsible for the mining procedure, which creates new blocks made by verified transactions. The mining procedure involves solving complex mathematical puzzles (PoW) or putting some currency on stake to validate the transactions (PoS).

The consensus mechanism employed by Ethereum has migrated from Proof of Work (POW) to Proof of Stake (PoS). The PoW used by the miners in the Bitcoin blockchain requires the miners to solve complex computational problems, creating new blocks for the blockchain. In contrast, the PoS mechanism selects validators based on the number of coins they stake. The network then chooses validators using probabilistic techniques to confirm the transactions and propose new blocks. Ethereum blockchain supports the creation of DApps for any organization supported by the smart contract at the application's backend. In the following section, we will explore smart contracts separately to identify their role in developing DApps.

**Overview of Smart Contracts**

Smart contracts are self-executing programs when certain defined conditions are met for the execution and written in solidity, Vyper and Web Assembly languages [34]. Solidity is most famous for writing smart contracts. The smart contracts are stored on the Ethereum blockchain as a computing entity and are an essential part of DApps as the functionality of the DApp is dependent on the smart contracts. Once deployed on the blockchain, smart contracts are immutable and tamper-resistant. Additionally, smart contracts eliminate the requirement of intermediaries and have features to communicate with other smart contracts and self-execution. The smart contracts provide secure and transparent transactions(operations) associated with the Ethereum blockchain [35].

**Development of Dapps using Ethereum Blockchain**:

DApps are blockchain-based applications that leverage decentralized databases instead of traditional centralized databases for storing and processing information. Meanwhile, fintech organizations typically use web applications backed by centralized databases. DApps use distributed file systems and decentralized networks(blockchain), enhancing security and transparency. Compared to traditional web applications, the DApss have more security due to their decentralized nature, making them resistant to single-point failure and unauthorized modifications. Support for decentralized application's security and transparency. The DApp operations are validated by the consensus mechanism, ensuring data integrity and trustworthiness. However, this verification process can increase the operation (transaction) processing time compared to traditional applications.

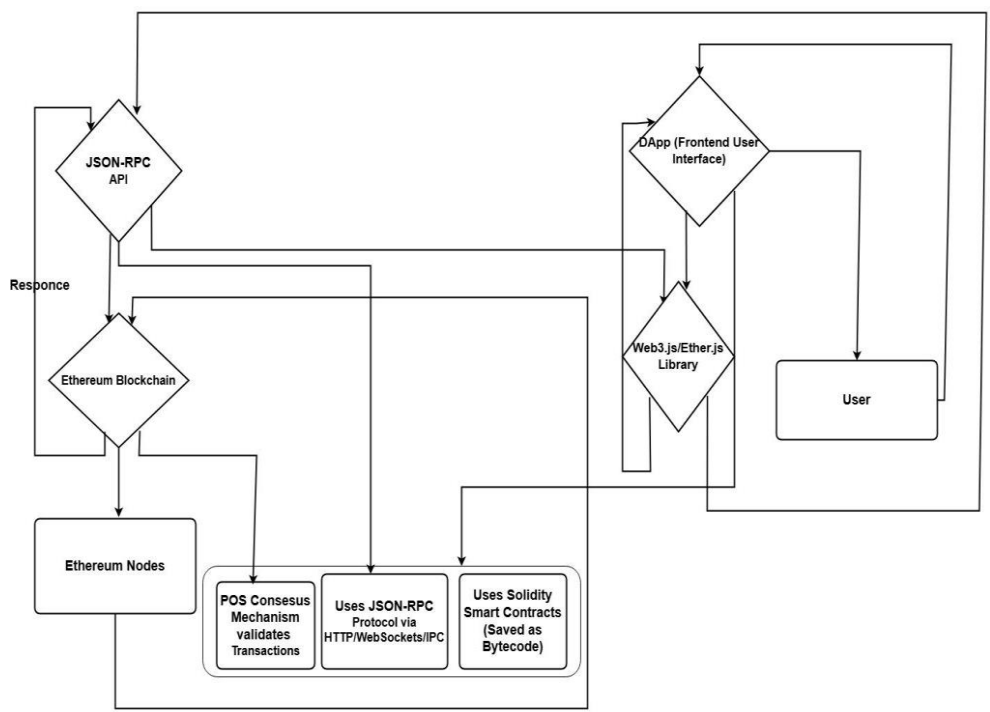

**Figure 3:** Decentralized Application Architecture and Components.

**Dapps have two components**

**i) The front-end user interface (DApp UI)** allows users to interact with the application, while the Web3.js or Ethers.js library provides communication between the front end and Ethereum blockchain via JSON-RPC API. The API transfers the data over HTTP, WebSocket or IPC protocols to the Ethereum nodes.

**ii) The back-end:** logic is implemented through smart contracts, primarily written in solidity, which are compiled using the solc compiler, and the outcome bytecode is deployed on the Ethereum blockchain at a given address.

When a user operates using DApp, the DApp sends a transaction request to the blockchain; the transaction is processed and validated using PoS consensus protocol. Once the transaction is validated, it is recorded on the blockchain, ensuring transparency, security and immutability. The Ethereum blockchain then responds to the valid request, updating the DApp with the latest state. This decentralized architecture of Dapp eliminates threats associated with a centralized structure, enhances security, and ensures trustless execution of operations. The DApp architecture with it's components are shown in Figure 3.

Given that a consensus protocol is crucial for the implementation of the security and integrity of the blockchain network, we will demonstrate the Proof of Stake protocol to identify the complete ecosystem of the blockchain network.

**Proof of Stake consensus protocol**

Proof of Stake consensus protocol stake (PoS) is a consensus mechanism employed by blockchain networks to validate transactions and add new blocks to the blockchain. Unlike the Proof of Work algorithm, the validation and verification of the transactions do not depend on the mining procedure but on the stake a validator applies in the blockchain network. The validator selection procedure depends on probabilistic techniques so that the transparency of the network remains fair, as well as decentralization and network transparency. The pseudocode of the algorithm for the validator selection is given in Algorithm 1 as follows:

**Algorithm 1: Validator selection algorithm**

Input:

S[] = Array of stakes for N validators R = Random number in [0, 1]

Output:

Selected validator ID

Step 1: Calculate total stake S_total = 0

for i = 1 to N:

S_total = S_total + S[i]

Step 2: Calculate probabilities

P[] = Array to store probabilities for i = 1 to N:

P[i] = S[i] / S_total

Step 3: Compute cumulative probabilities

C[] = Array to store cumulative probabilities C[1] = P[1]

for i = 2 to N:

C[i] = C[i-1] + Pi]

Step 4: Select validator for i = 1 to N:

if R <= C[i]:

return i // Validator i is selected

End

    The validator validates the transactions generated during DApp usage. Hence, the common consensus achieved by the blockchain network helps validate and add the transactions that are taking place while using a smart contract by the application or smart contract to smart contract communication.

    The DApp created for FinTech can use the above-mentioned mechanism to

perform transactions and other FinTech operations. In contrast, the computation is performed on the EVM, and transactions (successful computation process) are permanently stored on the blockchain. These transactions are immutable, transparent and secure, cooperating the security of the FinTech ecosystem.

**Proposed framework**

To understand the proposed framework using DApp, we must analyze the existing FinTech system used by a FinTech organization and how internal users are validated and verified based on a perimeter-based security paradigm. Next, we will identify the existing attack surface by creating a threat modeling diagram using the STRIDE threat model and OWASP Threat Modeler software to access vulnerabilities in the current Perimeter-based defence system [36]. The threat modeling diagram in Figure 4 presents the result of this analysis, highlighting the identified threats within the FinTech organization.

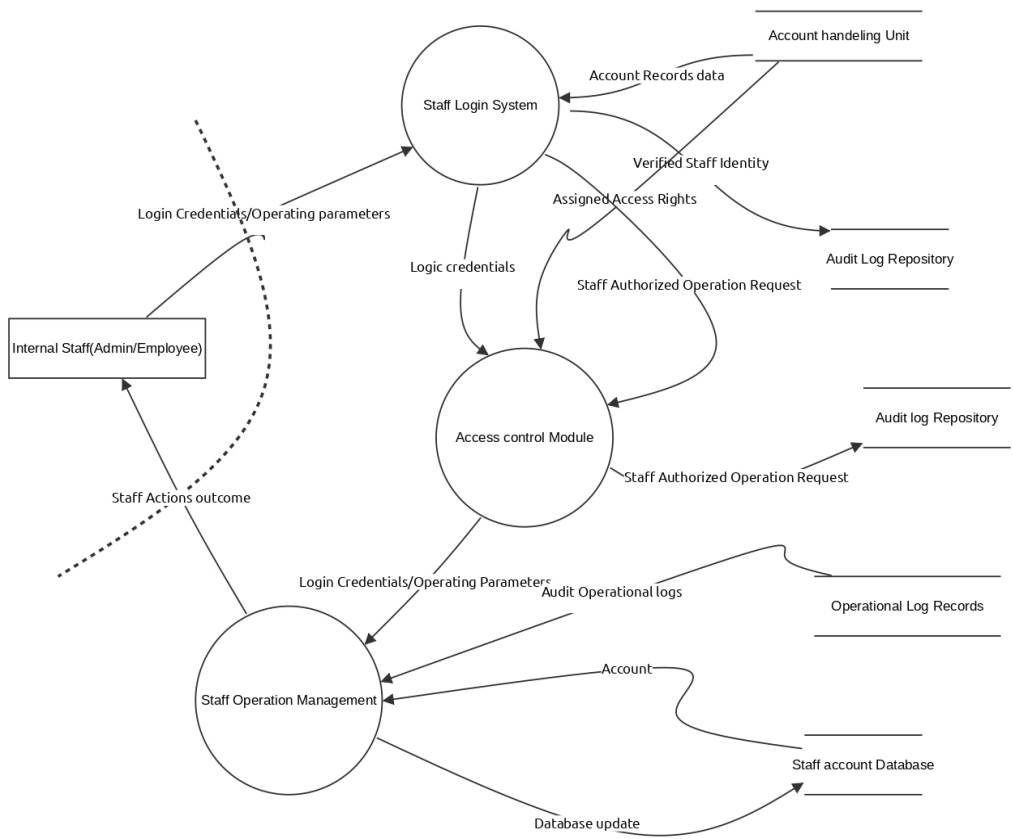

**Figure 4:** Threat model diagram for a fintech organization's internal working.

The above diagram includes different components, including Internal Staff as an entity and multiple processes, including the Staff Login System, Access Control Module, and Staff Operation Management. Additionally, it includes storage units such as the Account Handling Unit, Audit Log Repository, Operational Log Records, and Staff Account Database. These components, in combination, create a system responsible for internal user login and other operational activities of the fintech system.

The Internal staff initiates the process by submitting login credentials to the Staff Login System, which interacts with the Account Handling Unit to verify the user's identity. Once verified credentials are passed to the Access Control Module, the user's access rights are checked. Staff Operation Management processes all authorized requests and executes the tasks based on the request, such as updating account information, retrieving data, and handling other queries related to staff actions. All logs generated by the Staff Operation Management module are recorded in the Operational Log Records, while staff account-related data is stored in the Staff Account Database. Audit logs of authorized operations are stored in the Audit Log Repository for security and compliance processes.

The threat analysis performed using STRIDE threat modeling is presented in Table 1. This table includes details such as Threat Description, Potential Threat Actors, Relevant Threat Vectors, Assets at Risk, and Mitigation control Strategies, as represented in Table 2.

**Table 2: Threat Identification and Mitigation Strategies for Access Control Vulnerabilities.**

| Threat ID | Threat Description | Potential Threat Actor | Relevant Threat Vectors | Assets at Risk | Mitigation Control Strategies |
|---|---|---|---|---|---|

| T-1 | **Spoofing Staff Login**: An attacker impersonates internal staff to gain unauthorized access. | External attacker, malicious insider | Stolen login credentials, phishing, brute force attacks | Staff login credentials, sensitive financial data | *Implement MFA for all staff logins.* |
|---|---|---|---|---|---|
| T-2 | **Tampering with Access Rights**: Unauthorized modification of access control settings or policies. | Malicious admin, insider | Exploiting insecure APIs or privileged roles | Access control rules, sensitive data | *Enforce RBAC or ABAC for access control.* |
| T-3 | **Information Disclosure in Audit Logs**: Logs expose sensitive information. (e.g., credentials, PII). | Insider, external attacker | Poor log management, log injection attacks | Audit log repository | *Mask sensitive information in logs.* *Encrypt logs at rest and in transit.* *Restrict access to logs via role-based permissions.* |
| T-4 | **Repudiation of Staff Operations**: Staff members deny performing specific actions. | Malicious staff | Lack of proper audit trails or tamper-proof logging | Operational logs, accountability mechanisms | *Use tamper-evident logs.* *Digitally sign logs.* *Enforce time-stamped records for all actions.* |
| T-5 | **Denial of Service (DoS) on Staff Login System**: Overloading the login module, preventing legitimate access. | External attacker, botnet | Flooding login endpoints with invalid requests | Staff login system, system availability | Deploy CAPTCHA or bot detection tools. Implement rate-limiting on login attempts. - Use DoS protection. |
| T-6 | **Elevation of Privilege via Access Control Module**: Exploiting vulnerabilities to escalate privileges. | Malicious admin, insider | Exploiting poor privilege validation or unpatched vulnerabilities | Sensitive modules, user accounts | Regularly patch the system. Perform privilege validation checks dynamically. *Use just-in-time access for administrative privileges.* |
| T-7 | **Tampering with Database Update Requests**: Unauthorized changes to staff account data. | Insider, external attacker | SQL injection, man-in-the-middle attacks | Staff account database | Use parameterized queries or ORM to prevent SQL injection. |

The risks associated with the Staff Login System include spoofing(T-1) and denial of service (T-5); this module is a critical entry point that requires strong authentication. Due to its role in managing access rights, the Access Control Module is susceptible to risks like tampering (T-2) and elevation of privilege (T-6). The Staff Operation Management is vulnerable to database tampering (T-7) and repudiation (T-4). The storage units, including the Audit Log Repository, Operational Log records, Staff Account Database and Account Handling Unit, face the risk of information disclosure (T-3) if the logs are mishandled.

Our proposed framework integrates blockchain technology and features of Zero Trust principles to mitigate the attacks listed in Table 1 and reduce the attack surface of a fintech organization. We will present the proposed methodology and analyze the various components of the system. We will reapply the STRIDE threat model to compare the traditional perimeter-based defense approach and our blockchain-based zero trust implementation.

Additionally, we will present a simulation of the work by demonstrating the React.js-based application interface developed by the user, The interface allows internal access to the services within a fin-tech organization. This section will also introduce a case study of the proposed framework, showcasing the functionality of the Dapp. The DApp has features similar to those in the existing system but has the credibility of the blockchain supported by the Zero Trust principle. The following section will provide a detailed case scenario to elaborate on Dapp's operations.

Let us consider we have a company named "SecureFinance "that must provide its employees with access to sensitive financial records like customer transaction histories, account balances, and compliance documents. However, SecureFinance is highly concerned about insider threats - malicious or negligent employees, malware having the ability to steal credentials from the system and access information they shouldn't, potentially leading to spoofing (T-1), denial of service (T-5), Tampering (T-2) and elevation of privilege (T-6), database tampering (T-7) and repudiation (T-4) as mentioned in our previous section of the study.

Let us Assume three characters in our proposed work named Alice, Bob, and Charlie are internal employees of the company "SecureFinance".They use a Decentralized Application (DApp) developed using React front end and two smart contracts - MultifactorAuthentication and JustInTimeAccess-to enhance security. Alice and Bob are legitimate users, and Charlie is malicious. The malware software (acting as a bad actor) can steal the user credentials from the system. The stolen credentials could enable APT attacks, allowing sensitive information from the Fintech organization to be mined. In the next section, we will explain the functionality and workflow of our DApp.

**Functionality and workflow**

**1. Employee Registration (Using the front end and MultifactorAuthentication Contract)**

When Alice and Bob join SecureFinance, their administrator uses DApp's front end (Figure 5) to initiate the registration process. The frontend collects :

- E-mail address
- Chosen password
- Critical device information, including
  ➢ Device location (latitude and longitude)
  ➢ Browser details
  ➢ IP-Adress

- ➢ Operating system information (type and version)
- Checksum value: SHA-256 value of the device information.
- MAC-Id

This data is gathered using various operating system APIs. The front end generates a SHA-256 checksum of the device information for verification.

The administrator registers the user and assigns them a specific blockchain-enabled working node. The following information is stored on the Ethereum blockchain via the smart contract's register function:

- ➢ User's email address
- ➢ Hashed password
- ➢ device information (Checksum value)
- ➢ MAC address

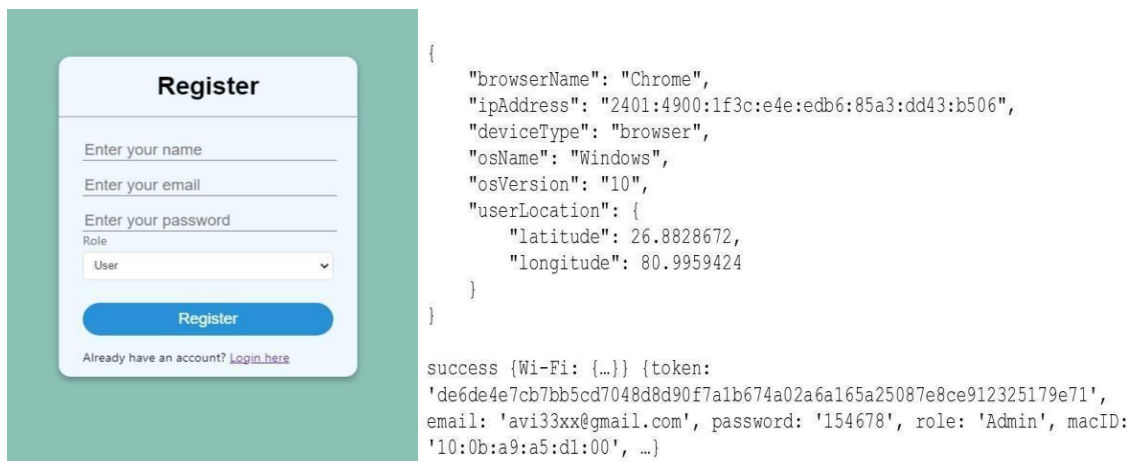

**Figure 5**: Registration page of Dapp.

These values are mapped in the smart contract to bind the user's identity to a specific Ethereum account address. This address serves as the user's unique identifier within the DApp. The two-phased registration process (front-end data collection and back-end recording) is completed by the administrator. The pseudocode of the register function is given below.

Data Structures:

User: Struct (userAddress, email, password, deviceInfo, macAddress, role, roleDescription)
users: Map (String (email) -> User) usersList: Array of Strings (emails)

State Variables: contractOwner: Address userCount: Unsigned Integer

Constructor:

contractOwner = msg.sender

Modifier: onlyOwner

IF msg.sender != contractOwner THEN

THROW ERROR: "Only the owner of the contract can perform this action."
ENDIF

Function: register(email, password, deviceinfo, macAddress) RETURNS Boolean
Input: email (String), password (String), deviceInfo (String), macAddress (String)
BEGIN
IF length(email) == 0 THEN
THROW ERROR: "name cannot be empty." ENDIF
IF length(password) == 0 THEN
THROW ERROR: "Password cannot be empty." ENDIF
IF length(deviceInfo) == 0 THEN
THROW ERROR: "deviceInfo cannot be empty." ENDIF
IF length(macAddress) == 0 THEN
THROW ERROR: "macAddress cannot be empty." ENDIF
IF users[name].userAddress != 0x0 THEN THROW ERROR: "Username already exists" ENDIF

userCount = userCount + 1

users[name] = User(msg.sender, emil, password, deviceInfo, macAddress, "NO_ROLE", "NO_DESCRIPTION")
usersList.push(name) RETURN true
END

**2. Role Assignment**

Following registration, the administrator assigns each user a unique role based on their organizational profile. Through the AssignRole function of the MultifactorAuthentication smart contract, the administrator can:
➢ Create new roles
➢ Define role description
➢ Assign roles to registered users

Concerning our use case scenario, if Alice has the Customer_Support role and its description includes ("Access to customer transaction histories") and if Bob is assigned the Compliance_Officer and role description to "Access the audit logs".

All roles and descriptions are permanently recorded on the blockchain, ensuring complete transparency and immutability. The pseudocode of the AssignRole function is given below:

Function: AssignRole(email, role, roleDescription) RETURNS Boolean

Input: email (String), role (String), roleDescription(String) BEGIN
IF length(role) == 0 THEN

THROW ERROR: "role cannot be empty." ENDIF
IF length(roleDescription) == 0 THEN

THROW ERROR: "roleDescription cannot be empty." ENDIF
IF users[email].userAddress == 0x0 THEN THROW ERROR: "User does not exist."
ENDIF

users[name].role = role users[name].roleDescription = roleDescription RETURN true
END

**3. Employee Login (using the front-end and MultifactorAuthentication smart contract)**

After registration, Alice initiates login by entering her email address and password. The frontend then :
Recalcultes her device information
Generates a new checksum( using the same SHA-256 algorithm as during registration)
calls the login function of the MultifactorAuthentication smart contract with:
➢ Email address
➢ Password (Securely hashed)
➢ Device information ( checksum value)
➢ MAC address

The system verifies all the above against the blockchain-stored data. If the all above matches with the stored data, the smart contract returns the active DApp page based on the user's role in using the services of the FinTech organization or returns an error according to the description. The login page of the DApp is shown in Figure 6, and the pseudocode for the login function is given below:

Function: login (email, password, deviceInfo, macAddress) RETURNS Boolean
Input: email (String), password (String), deviceInfo (String), macAddress (String)
BEGIN

```
IF length(name) == 0 THEN
THROW ERROR: "name cannot be empty." ENDIF
IF length(password) == 0 THEN
THROW ERROR: "Password cannot be empty." ENDIF
IF length(deviceLocation) == 0 THEN
THROW ERROR: "deviceLocation cannot be empty." ENDIF
IF length(macAddress) == 0 THEN
THROW ERROR: "macAddress cannot be empty." ENDIF
user = users[name]

IF user.userAddress == 0x0 THEN
THROW ERROR: "User does not exist." ENDIF
IF keccak256(user.role) == keccak256("NO_ROLE") THEN THROW ERROR: "Role is not assigned."
ENDIF
IF keccak256(user.password) != keccak256(password) THEN THROW ERROR: "Invalid password."
ENDIF
IF keccak256(user.deviceLocation) != keccak256(deviceLocation) THEN
THROW ERROR: "Invalid device location." ENDIF
IF keccak256(user.macAddress) != keccak256(macAddress) THEN THROW ERROR: "Invalid MAC address."
ENDIF
IF user.userAddress != msg.sender THEN THROW ERROR: "This is not your account"
ENDIF

RETURN true END
```

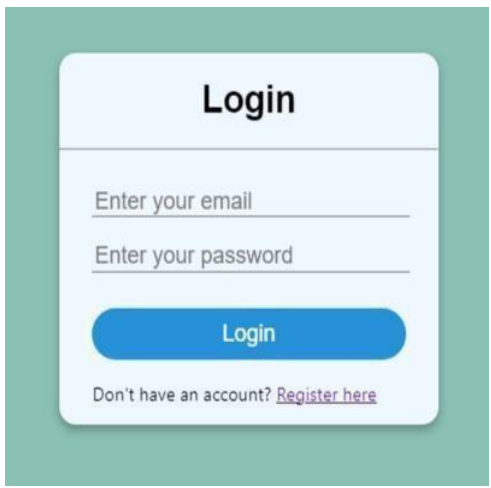

**Figure 6**: Login page of Dapp.

The implementation of the DApp also includes administrative privileges, allowing authorized admins to :
➢ Access all employee information within the FinTech organization.
➢ To get the total number of employees working in the Fintech organization

This function enables comprehensive user management while maintaining blockchain-based security. The pseudocode implementation of the getAllUsers function appears below:

Function: getAllUsers() RETURNS Array of User BEGIN
allUsers = new Array of User[length(usersList)]

FOR i = 0 TO length(usersList) - 1 DO allUsers[i] = users[usersList[i]]
ENDFOR
RETURN allUsers END

**4. Just-in-Time Access:**

In addition to the previously mentioned security measures, we implement a JustInTimeAccess smart contract to enforce the temporal restriction on the smart contract execution, including the MultiFactorAuthentication contract.

Consider the operational scenario: Bob (the compliance officer) needs to execute an "audit report" smart contract to generate a sensitive financial report. The process will work as follows:

1. The Dapp's frontend interacts with the JustInTimeAccess contract, providing
   - ➢ The target smart contracts address using Event-Driven control (audit report contract).
   - ➢ Required execution parameters.

2. The start execution function:
   - ➢ Records the initiation timestamp.
   - ➢ Sets an allowable execution window (Start time +threshold time).

3. Execution Control:
   - ➢ Bob can only execute the audit report smart contract within the approved time frame
   - ➢ The isOverTime function monitors the compliance:
   - ● Returns true if the execution is attempted outside the approved window.
   - ● Automatically terminates the unauthorized execution attempts.
.

This contract architecture ensures precise control over sensitive operations while maintaining the decentralized nature of fintech operations. The pseudocode implementation of the contract follows below:

Data Structure:

executionTime: Map (Contract Address -**>** Timestamp)

Function: startExecution(contractAddress) Input: contractAddress (Address)
BEGIN

IF contractAddress == 0x0 THEN\

THROW ERROR: "Invalid contract address" ENDIF
executionTime[contractAddress] = current_block_timestamp + Threshold_Value

END

Function: isOvertime(contractAddress) RETURNS Boolean Input: contractAddress (Address)
BEGIN

IF contractAddress == 0x0 THEN RETURN false
ENDIF

IF current_block_timestamp **>** executionTime[contractAddress] THEN RETURN true
ELSE

RETURN false ENDIF

END

Function: terminateExecution(contractAddress) Input: contractAddress (Address)
BEGIN

IF contractAddress == 0x0 THEN

THROW ERROR: "Invalid contract address" ENDIF
IF NOT isOvertime(contractAddress) THEN RETURN // Do nothing
ENDIF

result = CALL contractAddress.terminate() IF NOT result.success THEN
THROW ERROR: "Failed to terminate contract execution" ENDIF
END

```
·················································|························|··········|····························
|         Solc version: 0.8.18                  · Optimizer enabled: false · Runs: 200 · Block limit: 30000000 gas |
·················································|························|··········|····························
| Methods                                                                                                          |
·················|·······················|·········|·········|·········|···········|············
| Contract       · Method                · Min     · Max     · Avg     · # calls   · usd (avg) |
·················|·······················|·········|·········|·········|···········|············
| JustInTimeAccess         · startExecution    ·    -    ·    -    ·  44313  ·     3     ·     -      |
·················|·······················|·········|·········|·········|···········|············
| JustInTimeAccess         · terminateExecution·    -    ·    -    ·  24194  ·     1     ·     -      |
·················|·······················|·········|·········|·········|···········|············
| MultifactorAuthentication · assignRole      ·    -    ·    -    ·  42220  ·     3     ·     -      |
·················|·······················|·········|·········|·········|···········|············
| MultifactorAuthentication · register        · 219332  · 253556  · 245956  ·     9     ·     -      |
·················|·······················|·········|·········|·········|···········|············
| Deployments                              ·         ·         ·         · % of limit ·           |
·················|·······················|·········|·········|·········|···········|············
| JustInTimeAccess                         ·    -    ·    -    · 419174  ·   1.4 %    ·     -      |
·················|·······················|·········|·········|·········|···········|············
| MultifactorAuthentication                ·    -    ·    -    · 2275063 ·   7.6 %    ·     -      |
·················|·······················|·········|·········|·········|···········|············
```

**Figure 7:** Gas fees associated with smart contracts.

The execution of every smart contract on EVM requires gas fees, as represented in Figure7. The smart contracts are tested on local test net using hardhat and Mocha testing framework [36]; the test result is mentioned in Figure 8 below, having defined test cases associated with the smart contracts, both MultifactorAuthentication and JustInTimeAccess:

```
PS C:\Users\DELL\myproject> npx hardhat test

  JustInTimeAccess
The contract execution time has not expired
    ✓ should start the execution of a contract
    ✓ should check if a contract has exceeded its execution time
    ✓ should terminate the execution of a contract

  MultifactorAuthentication
=== 0xf39Fd6e51aad88F6F4ce6aB8827279cfffb92266
--- 0xf39Fd6e51aad88F6F4ce6aB8827279cfffb92266
    ✓ should register a new user
    ✓ should assign a role to a user
    ✓ should login a user
    ✓ should get all users
    ✓ should not allow duplicate user registration
    ✓ should not allow assigning a role by non-owner
    ✓ should not allow login with incorrect password
    ✓ should return an empty list of users if none are registered

  11 passing (13s)
```

**Figure 8**: Test report of smart contracts used for framework using Mocha framework.

In our case, we will consider an attack scenario where Charlie, a disgruntled employee, attempts to access the internal system. Charlie somehow obtains Alice's username and password (e.g. through malware) Charlie then tries to log into the DAPP using Alice's credentials from home or the other device is used in the fintech organization. The front end collects the information, which is different from Alice's device information because any change in the device information, including location, MAC address, browser information, Operating System information, and IP address, will reject the login procedure. The login function will return false, and Charlie will be denied access, preventing him from accessing sensitive customer data. The above-proposed framework will fulfil the following requirements of zero trust principles to prevent internal attacks in the fintech organization. The analysis of the zero trust principles is given below:

**Verify Explicitly (Multi-Factor Authentication)**

The DApp requires more than just a password to log in. It verifies the user's identity, device location, MAC address, and other information, including OS version, Browser information, and IP Address, making it much harder for attackers with stolen credentials to gain access.

**Least Privilege Access**

a. **Role-Based Access Control:** By assigning roles, the DApp ensures that employees only have access to the information and functionalities necessary for their job roles. Alice (customer service) won't have access to Bob's (compliance) audit tools, and vice versa.

**b. Just-In-Time Access:** The JustInTimeAccess contract limits the window of opportunity for malicious actions, even for authorized users. Sensitive operations are only allowed within a specific timeframe.

**Assume Breach (Device Verification)**

Even if an attacker compromises a user's password, the device verification acts as an additional barrier, limiting the damage they can do as they would need to be on a registered device. The checksum mechanism on the front end further strengthens this.

**Micro-segmentation**

The micro-segmentation among different devices is generated by the DApp itself and follows the segmentation by the blockchain; the DApp collects the information and, for unique user and device identity, will create an autonomous system separating it from the other systems. If a user is using a device, he cannot log in to other devices with different locations, device types, MAC IDs, IP addresses, and other parameters associated with a user device combo. The blockchain-enabled DApp creates the software-defined microsegments.

**Secure data and encrypt everything**

The smart contracts and valid transactions for access to the fintech resources are stored on the blockchain in encrypted form and are immutable. The communication between the DApp and Smart contract involves no encryption, and thus, this principle is partially addressed by the proposed framework and can be solved by using Zero-Knowledge Proofs (ZKP) or TLS protocols for secure communication.

**Secure all endpoints and devices**

This tenet's requirement can be manually fulfilled by the fintech organization's security team to update the operating system, browsers and other supportive software and generate a new checksum to store the device information on the blockchain.

**Continuous monitoring, automated security and response principles**

This goal can be achieved using AI-driven tools and techniques outside our framework's scope since our work is based on threat modeling analysis and mitigating the threats using the Zero Trust principal framework mentioned in Table 2. If implemented, these principles will enhance the organization's security posture.

**Transparency and Auditability (Blockchain)**

All registration, role assignments, and login attempts are recorded on the blockchain and verified by the consensus mechanism, providing an immutable audit trail. The decentralized application rejects suspicious activity; only valid activities are stored on the blockchain and verified by the consensus mechanism.

Now, we will perform a threat modeling approach to compare the existing and proposed systems in Table-3.

Table 3: Threat modelling report using blockchain and zero-trust

| Threat ID | Threat Description | Mitigation Control Strategies | The strategy implemented in the proposed framework using Zero-Trust |
|---|---|---|---|
| T-1 | **Spoofing Staff Login**: An attacker impersonates internal staff to gain unauthorized access. | - Implement the MFA for all staff logins. | MFA -Implemented |
| T-2 | **Tampering with Access Rights**: Unauthorized modification of access control settings or policies. | - Enforce RBAC or ABAC for access control. | RBAC -Implemented, and no tampering is possible on smart contracts stored on the blockchain. |
| T-3 | **Information Disclosure in Audit Logs**: Logs expose sensitive information (e.g., credentials, PII). | - Mask sensitive information in logs. - Encrypt logs at rest and in transit. - Restrict access to logs via role-based permissions. | RBAC -Implemented |
| T-4 | **Repudiation of Staff Operations**: Staff members deny performing specific actions. | - Use tamper-evident logs. Digitally sign logs. - Enforce time-stamped records for all actions. | The blockchain itself is tamper-evident, and valid logs are stored there. |
| T-5 | **Denial of Service (DoS) on Staff Login System**: Overloading the login module, | - Deploy CAPTCHA or bot detection tools. - Implement rate- | It is not applicable in decentralized networks like blockchain. |
| | preventing legitimate access. | limiting on login attempts. - Use DoS protection. | |
| T-6 | **Elevation of Privilege via Access Control Module**: Exploiting vulnerabilities to escalate privileges. | - Regularly patch the system. - Perform privilege validation checks dynamically. - Use just-in-time access for administrative privileges. | Just-in-Time administrative privileges are implanted using the JustInTimeAccess smart contract. |
| T-7 | **Tampering with Database Update Requests**: Unauthorized changes to staff account data. | Use parameterized queries or ORM to prevent SQL injection. | Blockchain itself tamper-evident no change is possible in the staff account data |

In the proposed framework, examined through a threat modeling approach, the blockchain functions as a Policy Engine(PE) and Policy Enforcement Point(PEP) and policy storage, as represented in Figure 2. All the access control activities are performed by the smart contracts, the policy rules encoded in smart contracts are stored on the blockchain. The Ethereum blockchain ecosystem, including DApps, smart contracts, and the Ethereum blockchain itself, fulfils the framework's requirements effectively. The blockchain enforces and verifies access rights within a fintech organization, ensuring that every internal communication is done through the blockchain. This approach reduces the attack surface and enhances the security concerning the perimeter-based security paradigm by limiting vulnerabilities and mitigating insider threats, including malware attacks stealing credentials and fintech organizations information and APTs due to blockchain-enabled Micro-segmentation. It also blocks SQL injections and data tampering due to the immutability of the blockchain. The proposed framework also denies the issue of single-point failure and DoS attacks because of the decentralized nature of the blockchain. The spoofing and elevation of privilege are countered by the framework's Multifactor Authentication (MFA) and Role-Based Access Control (RBAC) methodologies.

We have also computed the proposed framework's latency (Figure 9) and throughput

(Figure 10) using simulation over 200 nodes to demonstrate the difference between the two approaches.

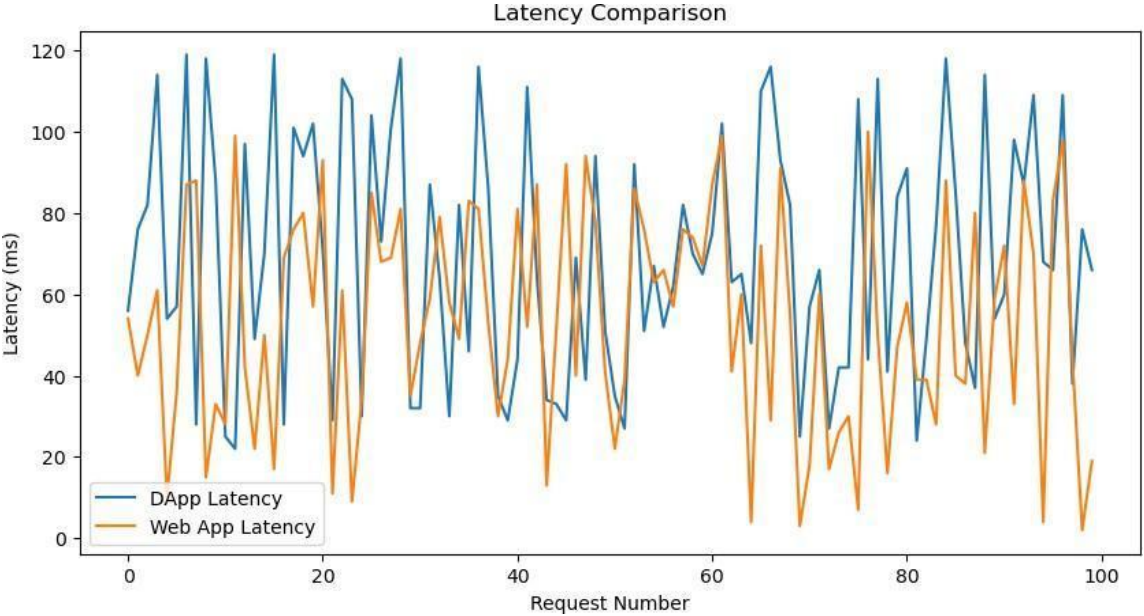

**Figure 9**: Latency comparison of the proposed approach with the traditional approach.

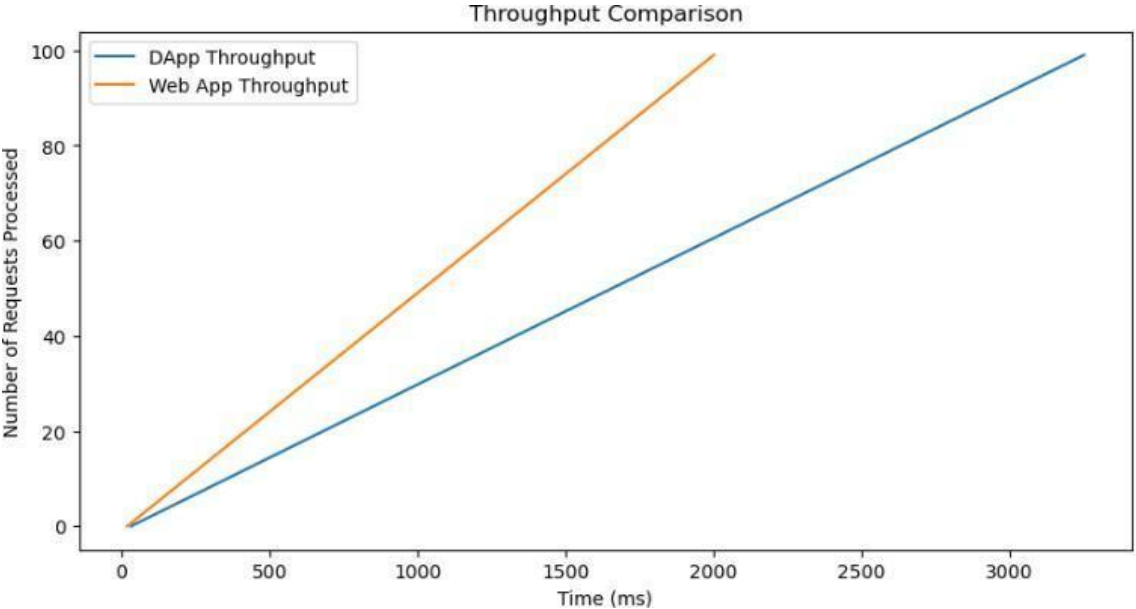

**Figure 10:** Throughput comparison of the proposed approach with the traditional approach.

The comparison of latency and throughput between the DApp and traditional web applications is given in Table- 4.

**Table 4: Latency and Throughput comparison of DApp vs Traditional web application.**

| Request Number | Depp's latency (milliseconds) | web application's latency (milliseconds) | DApp Time(Average time taken per request in milliseconds) | web applications Time(Average time taken per request in milliseconds) |
|---|---|---|---|---|
| 1 | 86 | 99 | 32.0 | 20.0 |
| 2 | 79 | 31 | 65.0 | 40.0 |
| 3 | 99 | 23 | 97.5 | 60.0 |
| 4 | 77 | 69 | 130.0 | 80.0 |
| 5 | 31 | 69 | 162.5 | 100.0 |
| 6 | 52 | 7 | 195.0 | 120.0 |
| 7 | 69 | 39 | 227.5 | 140.0 |
| 8 | 63 | 38 | 260.0 | 160.0 |
| 9 | 27 | 83 | 292.5 | 180.0 |
| 10 | 62 | 58 | 325.0 | 200.0 |

The average latency of the proposed framework-based DApp is 74.0 milliseconds, while the web application-based system operating on traditional perimeter-based defense has an average latency of 49.33 milliseconds. The Proposed framework exhibits higher latency values due to consensus overhead inherent in the blockchain-based system.

Additionally, the average throughput of the proposed framework-based DApp is 30.77 requests per second, whereas the web application-based system archives an average throughput of 50.0 requests per second. This indicates that traditional web application-based systems have a higher throughput and do not require a consensus mechanism for transaction validation.

**Result and Discussion**

**Result**

The experimental validation of the proposed blockchain-based Zero Trust framework demonstrates its effectiveness in mitigating security vulnerabilities while addressing the shortcomings of traditional perimeter-based defense mechanisms. The results are analyzed based on three key metrics:
- Security Resilience
- Performance overhead
- comparative analysis with the existing system

1. **Security Resilience:**

The STRIDE threat modeling analysis confirms that the proposed framework has the capability to reduce the attack surface in FinTech organizations. Key findings include:

**Mitigation of Insider Threats:** The integration of MultiFactor Authentication(MFA) and Role-Based Access Control (RBAC) created through smart contracts prevents unauthorized access, even when the credentials are compromised (e.g. malware or malicious insiders). The spoofing attempts(T-1) are precluded by the device-binding checks, while privilege escalation(T-6) was eliminated through Just-In-Time Access controls.

**Tamper-Proof Operations:** Blockchain's immutability ensures that the access policies and logs cannot be altered, addressing threats like data tampering (T-7) and repudiation (T-4). All transactions are cryptographically verified and stored on the blockchain, providing an auditable trail.

**Micro-Segmentation:** The binding of the user with a device with device attributes creates a DApp-driven segmentation, restricting the lateral movement of attackers within the network.

2. **Performance Overhead:**

While the framework enhances security, it adds measurable latency and throughput trade-offs due to the blockchain's consensus mechanism:

**Latency:** The average authentication latency for the proposed framework was 74.0 milliseconds compared to 49.33 for traditional web applications (Table 4). This increased value is due to smart contract execution and validation across nodes using consensus protocols.

**Throughput:** The DApp processed 30.77 requests per second, whereas the traditional web application achieved 50.0 requests per second (Figure 10). The lower throughput is due to the PoS consensus protocol's computational overhead.

Despite tradeoffs, the performance impact is justified by the security gains, particularly for high-stakes FinTech operations where data integrity and access control are critical.

3. **Comparative Analysis**

The proposed framework was evaluated against perimeter-based security models using STRIDE threat modeling (Table 3). Key advantages include:

**Elimination of Single Point of Failure**: Decentralization removes reliance on centralized authentication servers, mitigating DoS risks (T-5).

**Dynamic Policy Enforcement:** Smart contracts automate access control, reducing human intervention and policy misconfigurations.

**Cost Efficiency:** Blockchain's open-source nature lowers operational costs compared to proprietary security solutions.

However, the framework's reliance on Ethereum incurs gas fees (Figure 7) and scalability challenges for large-scale deployments. Future optimizations could explore Layer-2 solutions to improve throughput.

**Discussion**

The results validated that blockchain can effectively function as a Policy Engine (PE), Policy Enforcement Point (PEP) and Policy storage in Zero Trust architecture. The framework's success lies in its alignment with Zero Trust's key tenets:

**Explicit Verification:** MFA and device-binding ensures rigorous authentication.

**Least Privilege Access**: RBAC and JIT access minimize the exposure.

**Assume Breach:** Immutable logs and micro-segmentation contain the possibility of threats.

While latency and throughput are higher than traditional systems, the trade-off is acceptable for FinTech applications prioritizing security over speed. The framework's modular design also allows the integration with emerging technologies. (e.g. ZKP for encryption).

**Conclusion and future work**

**Conclusion**

This research demonstrates that integrating blockchain technology with Zero Trust principles offers a robust security framework for FinTech organizations, effectively mitigating insider threats, credential thefts, malware attacks, and APT attacks. By leveraging Ethereum smart contract to enforce Multi-Factor-Authentication (MFA), Role-Based-Access Control (RBAC) and Just-In-Time (JIT) access, the proposed framework addresses the shortcomings of perimeter-based defenses. Key outcomes include:

  **1. Enhanced Security:** The framework neutralized all STRIDE-identified threats (T-1 to T-7), mainly spoofing and privilege escalation, through immutable policy enforcement and micro-segmentation.

  **2. Decentralized Trust**: Blockchain's tamper-proof ledger eliminated a single point of failure and provided transparent audibility.

  **3. Practical viability:** Despite 24.67 milliseconds increase in latency and reduced throughput (30.77 vs 50.0 requests per second) the trade-offs are justified for high-security operations where data integrity outweighs speed.

**Future Work**

To optimize and extend the framework, future research should focus on:

**1. Performance Optimization:**

  ∗ Explore Layer-2 solutions (e.g. rollups, sidechains) to reduce Ethereum's gas fees and latency.

  ∗ Benchmark alternative blockchains (e.g. Hyperledger Fabric, Plkadot) for enterprise-scale deployments.

**2. Enhanced Security Features:**

  ∗ Integrate Zero-Knowledge Proofs (ZKPs) to secure frontend-backend communication, addressing partial encryption gaps.

  ∗ Implement AI-driven anomaly detection for real-time threat response, augmenting continuous monitoring.

**3. Broader Applications:**

  ∗ Improvements and additional feature development in the framework make it possible to use it for External users and mobile FinTech.

  ∗ Extend to multi-cloud environments where Zero Trust and blockchain can secure hybrid infrastructure.

  This study lays the groundwork for decentralized, trustless FinTech architectures, bridging the gaps between theoretical Zero Trust models and practical blockchain implementation.